\documentclass{article}

     \usepackage[final,nonatbib]{tccml_neurips_2020}

\usepackage[utf8]{inputenc} 
\usepackage[T1]{fontenc}    
\usepackage{hyperref}       
\usepackage{url}            
\usepackage{booktabs}       
\usepackage{amsfonts}       
\usepackage{nicefrac}       
\usepackage{microtype}      
\usepackage{chemformula}

\title{Deep Reinforcement Learning in Electricity Generation Investment for the Minimization of Long-Term Carbon Emissions and Electricity Costs}

\author{%
 Alexander J. M. Kell\thanks{Department of Chemical Engineering, Imperial College London, United Kingdom (This work was carried out whilst at Newcastle University).}\\
 \texttt{a.kell@imperial.ac.uk} \\
\And
Pablo Salas\thanks{Cambridge Centre for Environment, Energy and Natural Resource Governance (C-EENRG), University of Cambridge, United Kingdom.}  \\
\texttt{pas80@cam.ac.uk}\\
\And
Jean-Francois Mercure\thanks{Department of Geography, University of Exeter, United Kingdom.}\\
\texttt{J.Mercure@exeter.ac.uk} \\
\And
Matthew Forshaw\thanks{School of Computing, Newcastle University, United Kingdom} \\
\texttt{matthew.forshaw@ncl.ac.uk} \\
\And
 A. Stephen McGough\footnotemark[4]\\
 \texttt{stephen.mcgough@ncl.ac.uk} \\
}

\begin{document}

\maketitle

\begin{abstract}

A change from a high-carbon emitting electricity power system to one based on renewables would aid in the mitigation of climate change. Decarbonization of the electricity grid would allow for low-carbon heating, cooling and transport. Investments in renewable energy must be made over a long time horizon to maximise return of investment of these long life power generators. Over these long time horizons, there exist multiple uncertainties, for example in future electricity demand and costs to consumers and investors. 

To mitigate for imperfect information of the future, we use the deep deterministic policy gradient (DDPG) deep reinforcement learning approach to optimize for a low-cost, low-carbon electricity supply using a modified version of the FTT:Power model. In this work, we model the UK and Ireland electricity markets. The DDPG algorithm is able to learn the optimum electricity mix through experience and achieves this between the years of 2017 and 2050. We find that a change from fossil fuels and nuclear power to renewables, based upon wind, solar and wave would provide a cheap and low-carbon alternative to fossil fuels.

\end{abstract}

\section{Introduction}
\label{sec:intro}

To prevent climate change, a movement from a high carbon electricity supply to a low-carbon system is required \cite{Kell2020}. Low carbon electricity supply will aid in the decarbonization of the automotive and heating sectors by allowing for low-carbon electricity to be used in place of oil and gas.

Renewable energy costs, such as solar and wind energy, have reduced over the last ten years, making them cost-competitive with fossil fuels. These price drops are projected to continue \cite{IEA2015}. The future cost of operating and capital cost of electricity generation and electricity demand, however, remain uncertain over the long-term future. These uncertainties are risks which investors must analyze while making long-term decisions.

In this paper, we use the deep deterministic policy gradient (DDPG) reinforcement learning algorithm to simulate the behaviour of investors over a 33-year horizon, between 2017 and 2050 using the FTT:Power model \cite{Hunt2016a}. FTT:Power is a global power systems model that uses logistic differential equations to simulate technology switching \cite{Mercure2012}. The model is parameterized and runs from 2007; however, the investment decisions began in 2017. We start in this year due to the prior parameterization of the FTT:Power model with historical data up until this time. We projected until 2050 because this is a common target for governments to reach zero carbon. The environment used was a modified version of the FTT:Power model. 

We modified the FTT:Power model to use the DDPG algorithm in place of the logistic differential equations to make investment decisions. In addition, we simulated two countries: the United Kingdom and Ireland. We choose these due to the wealth of prior work on these countries which can be used use for comparison \cite{Hall2016, Hughes2010}. The DDPG algorithm allows us to simulate the decisions made by investors under imperfect information, such as future electricity costs, taxes and demand. This work enabled us to see an optimal, final state electricity generation mix.


Prior work in this domain has tackled the capacity expansion problem. For example, Oliveira \textit{et al.} also use reinforcement learning for the capacity expansion problem~\cite{Oliveira2018}. Whilst Oliveria \textit{et al.} provide detailed calculations of agents for the capacity expansion problem, we reduce this complexity to a series of observations of the environment, to allow for emergent behaviour. Kazempour \textit{et al.} use a mixed-integer linear programming approach to solve the generation investment problem \cite{Kazempour2011}. In contrast our approach removes the requirement for full knowledge of the time-horizon. 

Through this work, it is possible to assess whether a low cost, low-carbon electricity mix is viable over the long-term using a deep reinforcement learning investment algorithm, as well as finding what this optimum mix should be. This work enables us to closely match the investment behaviour of rational agents, without knowledge of the future. It can help guide investors on the choice and proportion of technologies to invest in over the long term.


\section{Model and methodology}
\label{sec:methods}

The Future Technology Transformations system for the power sector model (FTT:Power) model represents global power systems based on market competition, induced technological change and natural resource use and depletion \cite{Mercure2012}. This technological change is dependent on previous cumulative investment \cite{Mercure2012}. The model uses a dynamic set of logistic differential equations for competition between technology options.

For this work, we modified the FTT:Power model to use the deep reinforcement learning investment algorithm, DDPG. That is, the DDPG algorithm was used to make the decision on size of investment for each technology. In addition, we reduced the model only to consider the countries of Ireland and the UK. This enables us to iterate through enough episodes for the reinforcement learning to converge to an optimal reward. With more time it would be possible to undertake this optimisation for the entire world.

\subsection*{Reinforcement Learning}

The investment decision-making process can be formulated as a Markov Decision Process (MDP) \cite{puterman2014markov}. In an MDP environment, an agent receives an observation about the state of their environment $s_t$, chooses an action $a_t$ and receives a reward $r_t$ as a consequence of their action and the resultant change on the environment. Solving an MDP consists of maximizing the cumulative reward over the lifetime of the agent.

For our simulation environment, the agent makes continuous investment decisions for each energy technology, in each region and each year, starting from 2017 until 2050. Technology switching is modelled using a pairwise comparison of flows of market shares of different electricity generation capacity. That is, how much capacity flows from one technology to another. 

The agent's observation space is a vector consisting of the electricity produced by each technology, total capacity, total \ch{CO2} emissions over the simulation, levelized cost of electricity (LCOE) both with and without taxes, cumulative investment in each technology, investment in new capacity, carrier prices by commodity, fuel costs and carbon costs.

The reward $r$ is defined as:
\begin{equation}
\label{eq:reward_function}
	r = -\left(1000\times\ch{CO2}_e + \frac{LCOE}{1000}\right),
\end{equation}
where $\ch{CO2}_e$ is equal to total \ch{CO2} emissions over the simulation. The LCOE is calculated without taxes and the scaling factors are used to place the $LCOE$ and $\ch{CO2}$ on the same scale. The reward was multiplied by -1 due to the reinforcement learning (RL) algorithm maximizing reward and our requirement to reduce both LCOE and \ch{CO2} emissions.

RL approaches have been used to solve MDP through a trial and error based approach \cite{Sutton2015}. Since the paper published by Deep Mind in 2013 \cite{Arulkumaran2017}, RL has been extended to incorporate Deep Reinforcement Learning (DRL). DRL exploits deep neural networks to overcome the problems of memory and computational complexity \cite{Arulkumaran2017}. 

\begin{figure}
\centering
\includegraphics[width=0.5\columnwidth]{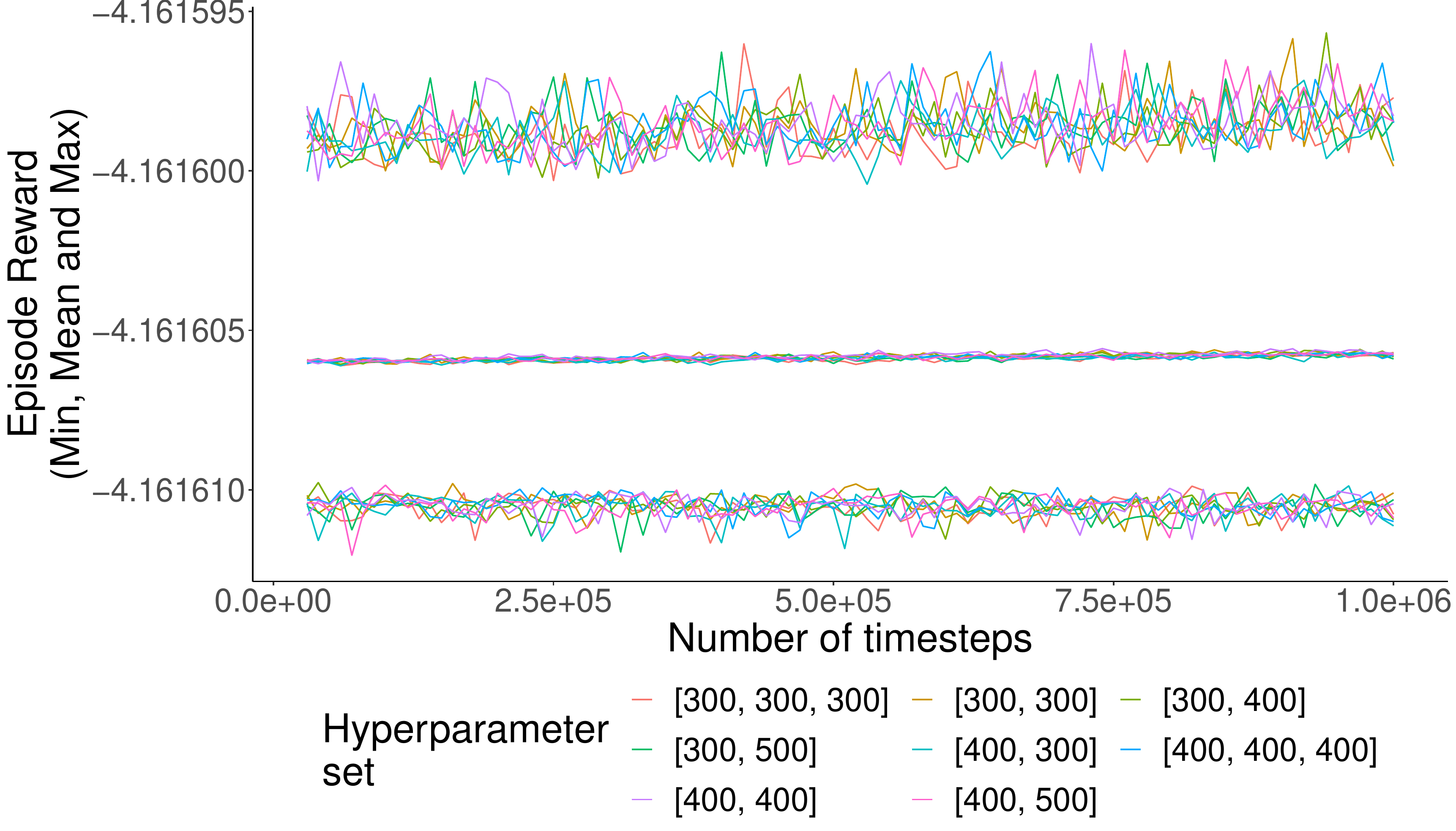}
\caption{Training with different hyperparameters, displaying the minimum, mean and maximum rewards per episode. The hyperparameter set [300, 500], for example, refers to two layers for both the actor and critic network, with 300 neurons in the first layers and 500 in the second.}
\label{fig:hyperparameter_training}
\end{figure}

We applied the deep deterministic policy gradient (DDPG) DRL algorithm \cite{Hunt2016a} from the Ray RLlib package to act as the investment algorithm \cite{Liang2014}. The DDPG algorithm is made up of an actor and critic network. We designed both of these to have two hidden layers, made up of 400 and 300 units per layer. The training batch size was set to 40,000. We chose these parameters as they were the default implementation in Ray RLlib. We trialled a variety of different configurations for the number of neurons per layer for hyperparameter tuning. To increase the speed of computation, just for the hyperparameter tuning, we reduced the simulation to run from 2007 to 2020. We chose this range as it allows for a change in the electricity mix. However, we found that the approach worked well, irrespective of parameter choice, as shown by Figure \ref{fig:hyperparameter_training}. 

%

\section{Results}
\label{sec:results}

%

\begin{figure}
\centering
\begin{minipage}{.4\textwidth}
  \centering
  \includegraphics[width=\linewidth]{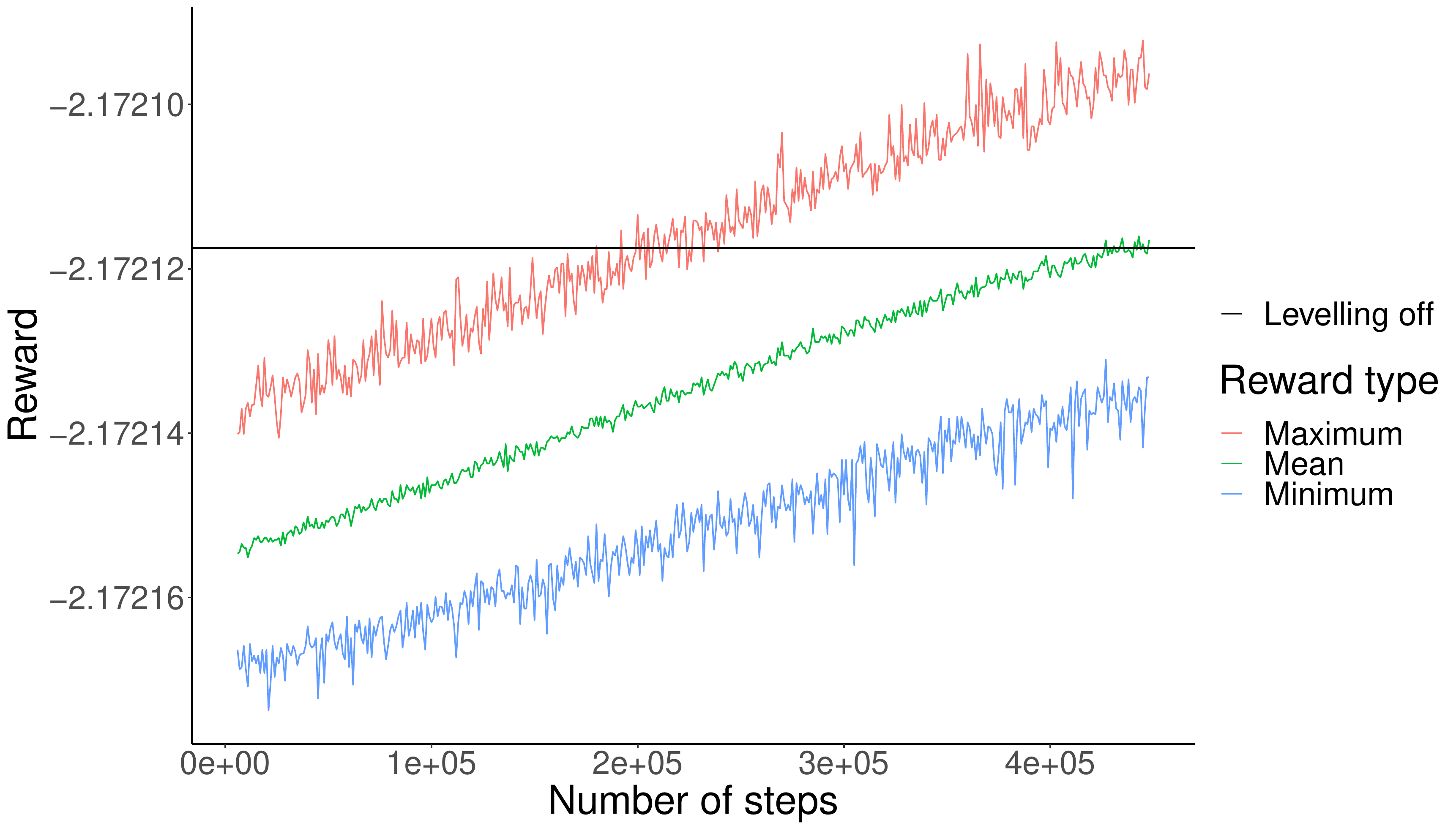}
  \caption{Mean, minimum and maximum rewards over run time.}
  \label{fig:days_reward_plot}
\end{minipage}%
\begin{minipage}{.4\textwidth}
  \centering
  \includegraphics[width=\linewidth]{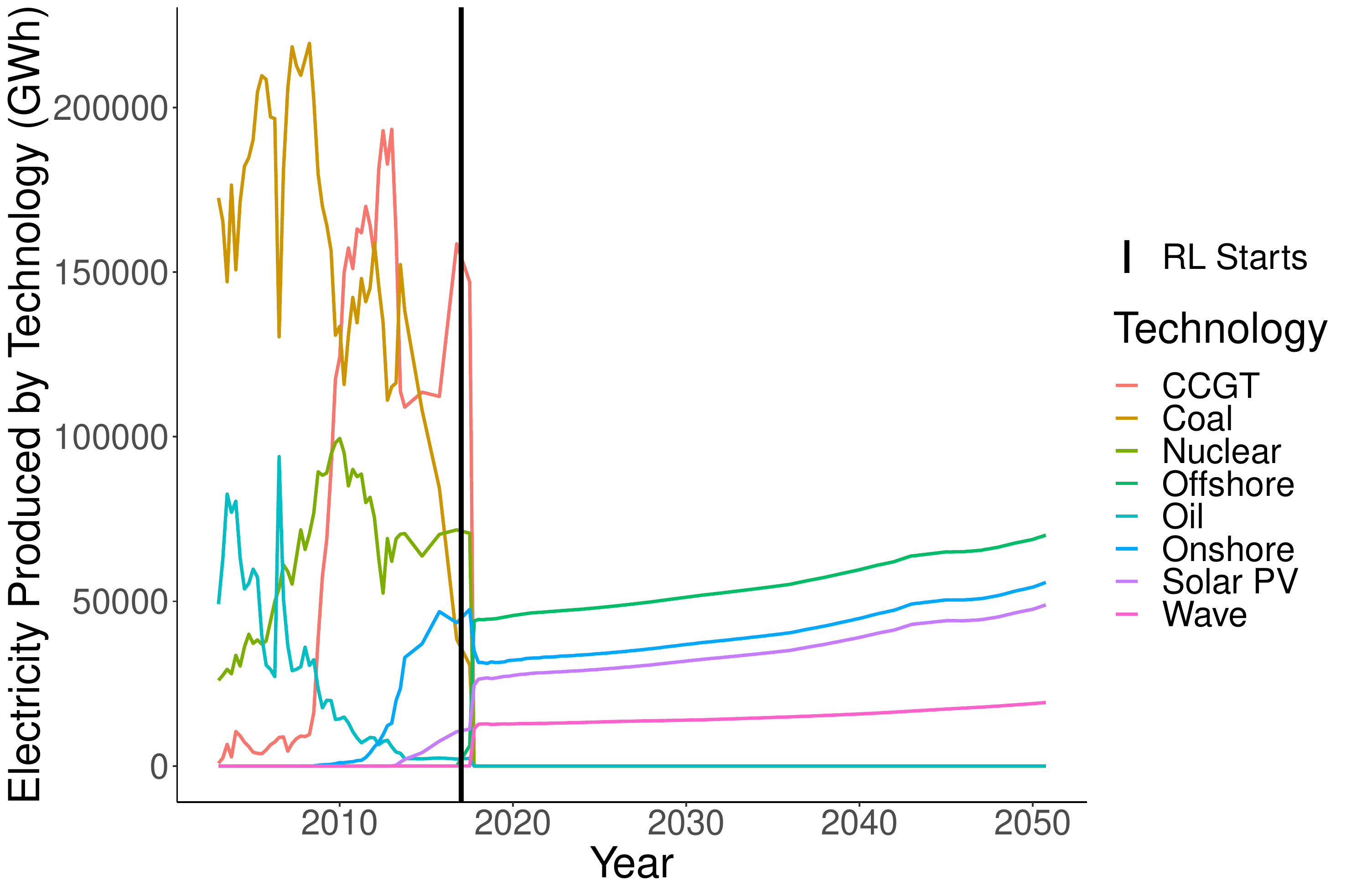}
  \caption{Electricity mix over time.}
  \label{fig:electricity_generated_plot}
\end{minipage}
\end{figure}

Our results show that our investment agent can increase its reward over time, as shown in Figure \ref{fig:days_reward_plot}. A total of ${\sim}$400,000 steps were required to see a levelling off in reward. The total time to simulate ${\sim}$400,000 steps was ${\sim}$8 days. We stopped the training and simulation after this time due to diminishing returns and the cost of computation.

Figure \ref{fig:electricity_generated_plot} displays the results of the reinforcement learning algorithm. Before the black vertical line (2017), the investments made are based upon historical data used by FTT:Power. The reinforcement learning algorithm starts to make investments after the black vertical line.

The historical electricity mix before 2017 is based mainly on fossil fuels: coal, combined cycle gas turbine (CCGT) and oil. Additionally, nuclear is a significant component of the electricity mix before 2009. After reinforcement learning optimizes for LCOE and carbon emissions, a rapid change occurs from fossil fuel and nuclear to renewable energy. 

This sudden	 change occurs because the RL algorithm does not take into account the technical and timeframe constraints embedded in the unmodified FTT:Power model. However, although it is likely that whilst the transition speed is unrealistic, the electricity mix found by the reinforcement learning algorithm is likely to be optimal, according to the reward function defined in Equation \ref{eq:reward_function}. We, therefore, show what the future should look like.


\begin{figure}
\centering
\begin{minipage}{.4\textwidth}
  \centering
  \includegraphics[width=\linewidth]{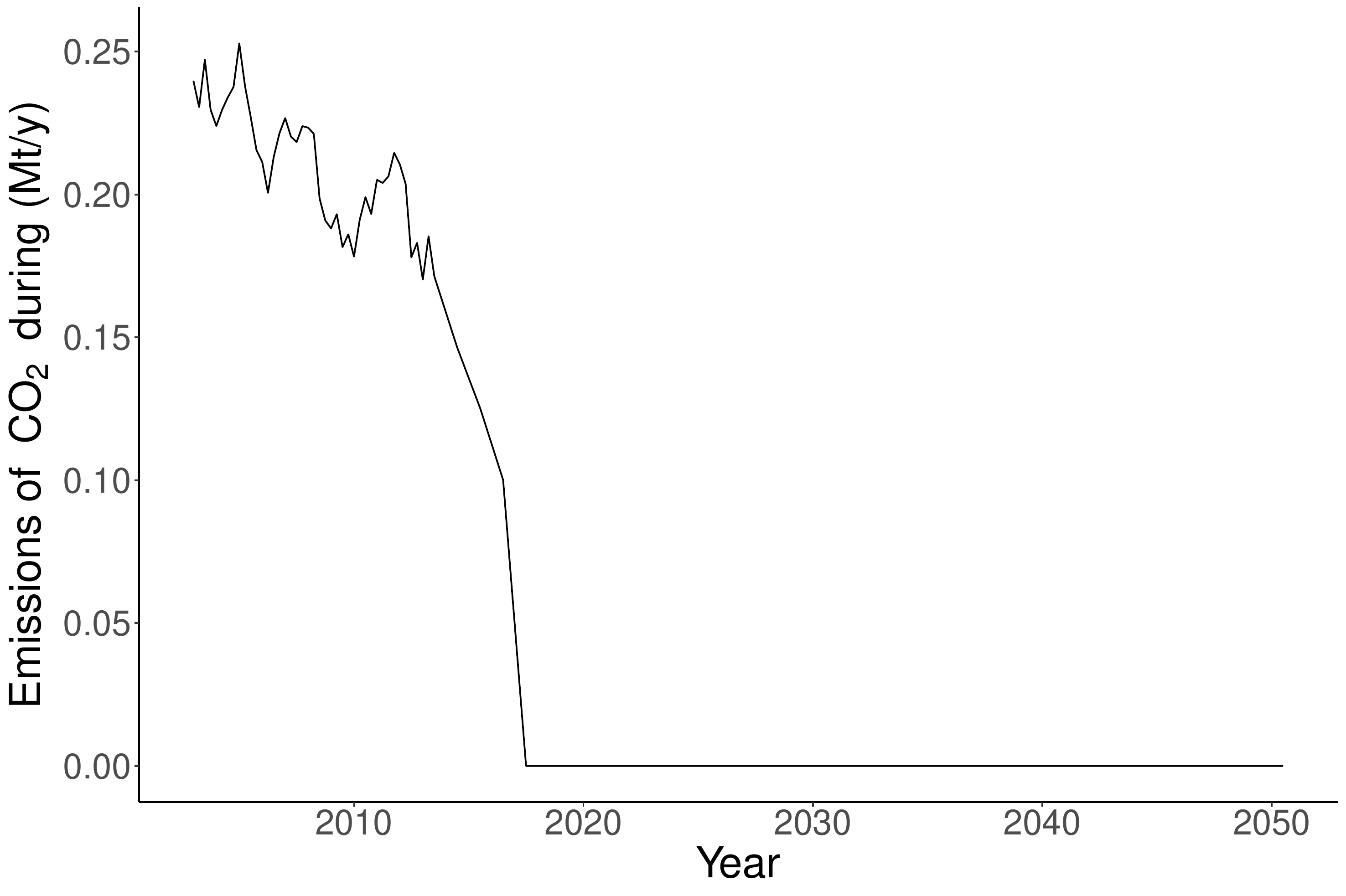}
  \caption{Carbon emissions.}
  \label{fig:emissions_plot}
\end{minipage}%
\begin{minipage}{.4\textwidth}
  \centering
  \includegraphics[width=\linewidth]{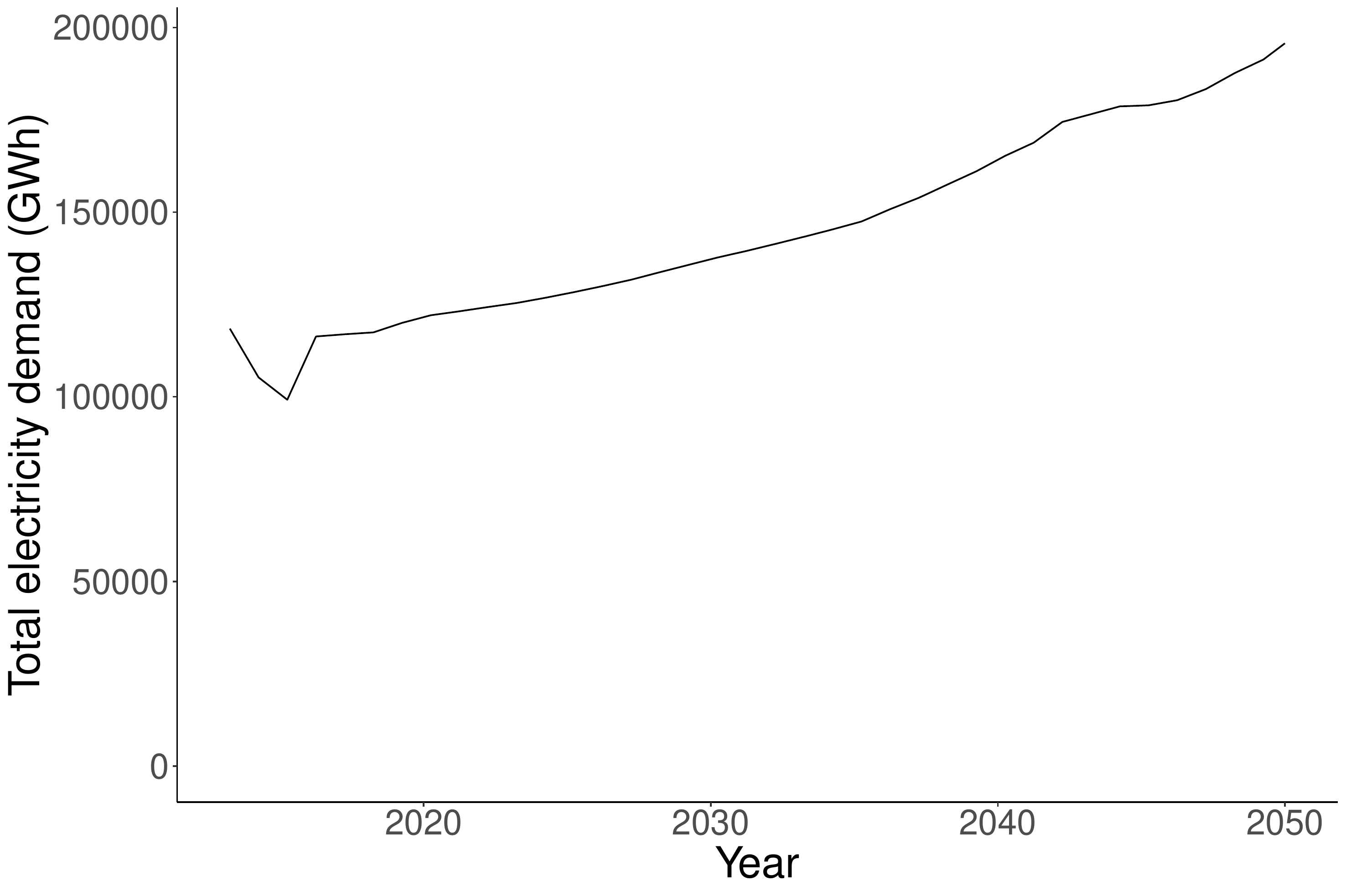}
  \caption{Demand scenario.}
  \label{fig:demand_scenario}
\end{minipage}
\end{figure}

The primary source of energy after the reinforcement learning algorithm begins is offshore, followed by onshore, solar photovoltaics (PV) and wave. As can be seen by Figure \ref{fig:emissions_plot}, the carbon emissions reduce significantly at the time that the reinforcement learning algorithm begins to control investments. 

This mix of renewable electricity generation across Ireland and the UK allows for demand to be met during the quarterly time periods of the model. The demand scenario is shown in Figure \ref{fig:demand_scenario}, where the demand can be seen to closely match the electricity mix shown by Figure \ref{fig:electricity_generated_plot}.

\section{Discussion}

A change from a high carbon-emitting electricity grid to a low-carbon system is required. In order to achieve this, investments in electricity generators must be made whilst taking into account future uncertainty. In this paper, we have modelled a central agent which makes investment decisions in an uncertain environment to find an optimal low-cost, low-carbon electricity mix. To achieve this, we used the reinforcement learning algorithm, DDPG. The environment is modelled using FTT:Power.

Through this exercise, we are able to see the optimal electricity mix in the UK and Ireland. We found that a mixture of renewable sources such as wind, solar and wave power would meet demand at quarter year intervals, as well as providing a cost-effective and low-carbon system.

A limitation of this work is the fact that the investment algorithm does not take into account the technical and timeframe constraints of transitions between technologies. It is for this reason that the reinforcement learning algorithm is able to make such a rapid change in 2017. However, we believe that the investment algorithm is able to find a general solution to the problem of investing in a cost-efficient and low-carbon system over a long time horizon. In future work, we would like to model the transition required by incorporating the technical and timeframe constraints for technology switching. This could be undertaken by modifying the reward function to ensure the transition remains within these constraints.

We would like to increase the number of steps of the FTT:Power model to more adequately model the investment behaviour introduced by the reinforcement learning algorithm. A lower number of simulated time steps leads to an overestimation of the supply of renewables and underestimation of storage and dispatchable technologies \cite{Ludig2011}. In addition, an increase in the number of countries modelled would enable us to see a global picture of how different, interdependent regions may evolve in a new climate of a requirement of low-carbon emissions. This would require an exponentially longer runtime for the reinforcement learning algorithm to converge. This is due to the increased number of decisions that the reinforcement learning algorithm would need to make to account for the different countries.

\section{Acknowledgements}

This work was supported by the Engineering and Physical Sci- ences Research Council, Centre for Doctoral Training in Cloud Computing for Big Data [grant number EP/L015358/1].

\bibliographystyle{ieeetr}
\bibliography{library,ftt-power-custom}

\end{document}